\documentclass[a4paper,12pt]{article}
\usepackage{jheppub}
\usepackage{graphicx}  
\usepackage{bm}        
\usepackage{amssymb}   
\usepackage{slashed} 
\usepackage{amsfonts}
\usepackage{wasysym}
\usepackage{amssymb}
\usepackage{hyperref}
\usepackage[applemac]{inputenc}
\usepackage{textcomp}
\usepackage[caption=false]{subfig}
\usepackage{braket}
\usepackage{placeins}
\usepackage{color,colortbl}
\usepackage[british]{babel}

 \newlength{\wth}
\newcommand{\be}{\begin{equation}}
\newcommand{\ee}{\end{equation}}
\newcommand{\bea}{\begin{eqnarray}}
\newcommand{\eea}{\end{eqnarray}}
\newcommand{\nn}{\nonumber}

\newcommand{\ti}{\times}

\newcommand{\beqa}{\begin{eqnarray}}
\newcommand{\eeqa}{\end{eqnarray}}

\newcommand{\lp}{\left(}
\newcommand{\rp}{\right)}

\newcommand{\ep}{\epsilon}

\renewcommand{\O}{\mathcal{O}}

\newcommand{\beq}{\begin{equation}}
\newcommand{\eeq}{\end{equation}}

\newcommand{\bW}{\ensuremath{\overline W}}
\newcommand{\bT}{\ensuremath{\overline T}}
\newcommand{\bF}{\ensuremath{\bar F}}
\newcommand{\bS}{\ensuremath{\bar S}}
\newcommand{\bZ}{\ensuremath{\overline Z}}

\newcommand{\bX}{\ensuremath{\overline X}}
\newcommand{\bb}{\ensuremath{\bar b}}
\newcommand{\ba}{\ensuremath{\bar a}}

\newcommand{\bi}{\ensuremath{\bar \imath}}
\newcommand{\bj}{\ensuremath{\bar \jmath}}
\newcommand{\bk}{\ensuremath{\bar k}}
\newcommand{\od}{\ensuremath{{\cal O}(\epsilon)}}
\newcommand{\ods}{\ensuremath{{\cal O}(\epsilon^2)}}
\newcommand{\odc}{\ensuremath{{\cal O}(\epsilon^3)}}
\newcommand{\odz}{\ensuremath{{\cal O}(\epsilon^0)}}

\usepackage{tikz}
\usetikzlibrary{fit}
\tikzset{%
  highlight/.style={rectangle,rounded corners,fill=red!15,draw,
    fill opacity=0.5,thick,inner sep=0pt}
}

\numberwithin{equation}{section}

\title{Decoupling and de Sitter Vacua in Approximate No-Scale Supergravities}

\author[1]{M.C.~David Marsh}
\author[2]{, Bert Vercnocke}
\author[3]{and Timm Wrase}
\affiliation[1]{Rudolf Peierls Centre for Theoretical Physics, University of Oxford,\\ 1 Keble Road, Oxford OX1 3NP, United Kingdom}
\affiliation[2]{Institute of Physics, University of Amsterdam, Science Park,\\
Postbus 94485, 1090 GL Amsterdam, The Netherlands}
\affiliation[3]{
Stanford Institute for Theoretical Physics, Department of Physics, Stanford University, 382 Via Pueblo Mall, Stanford, CA 94305-4060, USA}
\emailAdd{david.marsh1@physics.ox.ac.uk}
\emailAdd{bert.vercnocke@uva.nl}
\emailAdd{timm.wrase@stanford.edu}

\abstract{
We study ${\cal N}=1$ supergravity with $N>1$ chiral superfields in which one of the fields has a K\"ahler potential of exact no-scale type. Such systems admit de Sitter (dS) solutions in which supersymmetry is predominantly broken by the no-scale field, with only a small contribution to the breaking coming from the other fields. Metastable dS vacua of this type were recently shown to be achievable by the finetuning of an $N\times N$ sub-matrix of the Hessian matrix at the critical point. 
We show that perturbatively small deformations of the no-scale Minkowski vacuum 
into dS are only possible when the spectrum of the no-scale vacuum, besides the no-scale field, contain an additional massless mode.
The no-scale structure allows for  a decoupling of $N-2$ fields, and  metastability can be achieved by the tuning of ${\cal O}(N^0)$ parameters.
We illustrate this scenario in several examples, and derive a geometric condition for its realisation in type IIB string theory. Supergravities in which the complex structure moduli space is a symmetric space, such as the string theory inspired STU-models, are non-generic and realise a modified version of the scenario. For the STU-model with a single non-perturbative correction we present an explicit analytic family of dS solutions
that includes examples with quantised fluxes satisfying the O3-plane tadpole condition.
}

\begin{document}

\makeatletter
\let\old@fpheader\@fpheader
\renewcommand{\@fpheader}{
\hfill SU/ITP-14/27}
\makeatother

\maketitle

\section{\label{sec:intro} Introduction}

The observation of an apparently accelerated expansion of the universe \cite{Riess:1998cb, Perlmutter:1998np} is perhaps the most striking discovery of modern cosmology. Recent, combined  data  from the Cosmic Microwave Background (CMB) and distant supernovae gives strong evidence for the  energy density of the universe being dominated by a cosmological constant \cite{Union2, Ade:2013zuv}, thus making the present universe well approximated by four dimensional de Sitter space.

String theory may accommodate a small positive cosmological constant through the numerous metastable solutions forming  the `landscape' of  type IIB flux vacua, but determining the statistical properties of such vacua has proven challenging. While constructing fully generic flux vacua involving a large number of moduli  is computationally prohibitive, much can be learned from subsets of solutions in which a substantial number of moduli can be made to {\it decouple}.  In the KKLT scenario \cite{Kachru:2003aw}, complex structure moduli are supersymmetrically stabilised at a scale that is much larger than the scale of supersymmetry breaking, $m_{3/2}$, thus ensuring that these fields generically can be integrated out of the Wilsonian effective field theory and do not develop tachyonic directions upon supersymmetry breaking.
In the Large Volume Scenario (LVS) \cite{Balasubramanian:2005zx, Conlon:2005ki}, complex structure moduli obtain masses of the order of $m_{3/2}$, but an underlying no-scale symmetry again leads to  decoupling and a substantially more tractable low-energy dynamics, at least in the non-supersymmetric AdS minimum.\footnote{
 For a recent discussion on  the spectrum of the complex structure moduli
in  de Sitter space, see \cite{Sousa:2014qza}.}  In both KKLT and LVS, the final metastable de Sitter solution may be obtained by the inclusion of some additional, `uplifting', source of supersymmetry breaking.

Recently, the construction of metastable de Sitter vacua through spontaneous supersymmetry breaking has received a lot of attention (see for example \cite{Danielsson:2011au, Rummel:2011cd, Cicoli:2012fh, Cicoli:2012vw, Hebecker:2012aw, Louis:2012nb, Danielsson:2012by, Blaback:2013ht, Damian:2013dq, Damian:2013dwa, Cicoli:2013cha, Blaback:2013qza, Danielsson:2013rza, Rummel:2014raa, Ferrara:2014kva, Kallosh:2014via, Kallosh:2014wsa} and references therein). By constructing solutions that are consistently captured by spontaneously broken ${\cal N}=1$ supergravity, one can hope for improved computability and a more detailed picture of this part of the `landscape'. Particularly interesting are those solutions which make use of an exact or approximate no-scale symmetry of the K\"ahler potential for some of the fields. No-scale K\"ahler potentials are commonly encountered in the dimensionally reduced, four dimensional  effective theories arising from compactifications of string theory, and are in the simplest case of a single no-scale modulus $T$ and $N-1$ other moduli $X^i$ with $i=1,\ldots N-1$, given by,
\be
K = K_{\rm no{\text -}scale}(T, \bT) + \tilde K(X^i, \bX^{\bi}) = -3 \ln \left( T + \bT\right) +\tilde K(X^i, \bX^{\bi})
\, ,
\label{eq:K}
\ee
while the superpotential is independent of $T$,
\be
W= W_0(X^i) \, .
\label{eq:W0}
\ee
The perhaps most well-known example of such an effective theory is the dimensionally reduced type IIB compactification with a single K\"ahler modulus $T$, $N-2$ complex structure moduli, $U^i$, and the axio-dilaton $S$.

The F-term supergravity potential is remarkably simplified for no-scale models,
\be
V = e^K\left( F_a \bF^a - 3 |W|^2 \right) = e^K F_i \bF^i \, ,
\label{eq:V0}
\ee
where $a$ runs over all fields, $i$ runs over the $X^i$ fields, $F_a= D_a W= (\partial_a +K_a)W$ and we have used natural units to set $M_{\rm Pl} = 2.4 \times 10^{18}\, $GeV to one.  For $F_i=0$, as  in the `GKP' type IIB flux compactifications of \cite{Giddings:2001yu}, the $X^i$ fields are supersymmetrically stabilised and $T$ remains a flat direction of the Minkowski vacuum.

Recently, motivated by the numerical solutions of \cite{Saltman:2004sn, Blaback:2013qza, Danielsson:2013rza}, reference \cite{Kallosh:2014oja} constructed a class of analytical de Sitter solutions in which supersymmetry is predominantly broken by the no-scale field $T$, with only a small amount of supersymmetry breaking in the perpendicular directions. For type IIB realisations of this mechanism, the smallness of the supersymmetry breaking in the directions perpendicular to $T$ can be achieved by tuning of fluxes, as discussed in \cite{Saltman:2004sn}.
A given critical point is a metastable minimum of the potential if all the eigenvalues of the $2N \times 2N$ hermitian Hessian matrix,
\be
{\cal H} =
\left(
\begin{array}{c c}
\partial^2_{a \bb} V & \partial^2_{a b} V \\
\partial^2_{\ba \bb} V & \partial^2_{\ba b} V
\end{array}
\right) \, ,
\label{eq:Hess}
\ee
are positive. In \cite{Kallosh:2014oja}, analytical conditions for the positivity of  the  eigenvalues of the
diagonal $N\times N$ sub-matrix  $\partial^2_{a \bb} V$ were derived, and it was shown that upon tuning
the $N\times N$ off-diagonal  block $\partial^2_{a b} V$ to vanish or to be very small, the full Hessian could be made positive definite.

This type of solutions were shown to be realisable in string theory inspired `STU' supergravity models with $N=3$, however, a general concern with this method is the significant fine-tuning needed to ensure metastability: for $N\gg1$, the required ${\cal O}(N^2)$ fine-tuning of $\partial^2_{a b} V$ can be greatly limiting.

In this paper, we show that, given a small amount of no-scale breaking, there are approximate no-scale solutions that require only minimal, ${\cal O}(N^0)$, tuning to ensure the metastability of the Hessian matrix.
To do this, we systematically expand the Hessian matrix, equation \eqref{eq:Hess}, in the small no-scale breaking parameter $\epsilon$. In particular we note that to zeroth order in no-scale breaking,  $\partial^2_{a b} V$ is not small, yet has a structure that ensures that the mass matrix for all $X^i$ fields is semi-positive definite. This is a well-known property of flux compactifications of GKP type, that we here fully explore.

No-scale Minkowski vacua have  at least  two real  massless fields  
corresponding to the flat directions along
the complex  field $T$. We find that, to zeroth order in $\ep$, the no-scale critical points that can be perturbed into de Sitter vacua
in addition
 always have a third real massless mode 
 in the spectrum of the fields $X^i$. This 
 additional flat direction
  is generically lifted at linear order $\epsilon$, while the complex no-scale field $T$ is  lifted at order $\epsilon^2$. The 
   underlying no-scale symmetry ensures that the 
   $2N-3$ real fields that are lifted at
    order ${\cal O}(\ep^0)$ are  metastable with positive definite masses.
A remarkable decoupling of the heavy modes in the Hessian matrix at the approximately no-scale critical points then ensure that metastability can be achieved by the tuning of only two terms, independently of $N$.

As an illustration, we contrast these findings to those obtained in  `random supergravity' in which the superpotential and K\"ahler potential are taken to be random functions in the sense of \cite{Marsh:2011aa}, and in which only a fraction $P \lesssim {\rm exp}( -N)$ of the critical points are metastable minima.\footnote{
Typical critical points in random supergravity for which the supersymmetry breaking scale of the same order as the supersymmetric masses,  $m_{3/2} \approx m_{susy}$,   the
fraction of metastable critical points, $P$,  scale with $N$ like $\ln P \sim -N^2$. 
For approximately supersymmetric critical points with $m_{3/2} \ll m_{susy}$, the corresponding fraction scales like $\ln P \sim -N$.} For the approximately no-scale critical points constructed in this paper, we show that the corresponding fraction  is $1/2$, independently of $N$ and the value of the gravitino mass.

The general mechanism presented in this paper appears readily embeddable in flux compactifications of type IIB string theory, and 
we 
derive a sufficient condition on the complex structure field space geometry for the realisation of the mechanism with minimal ${\cal O}(N^0)$ fine-tuning.
When the complex structure field space is a symmetric space, the condition is violated and 
${\cal O}(N)$ entries of the Hessian matrix need to be tuned  to ensure stability. 
However, even when the geometric condition is not satisfied, 
approximate no-scale dS vacua can be obtained with a moderate tuning, as  we explicitly illustrate
 by constructing simple de Sitter solutions in STU supergravity.

While the realisations of this supersymmetry breaking scheme in the simplest models with a single no-scale field are somewhat restricted and do not allow for an exponentially large volume, we know of no reason why the general mechanism could not be applied to the case of multiple K\"ahler moduli in which an exponentially large volume might be achievable.
Extending this scheme to more general sectors of no-scale fields is an important future direction.

This paper is organised as follows: In \S\ref{sec:noscale}, we review no-scale Minkowski vacua of $\mathcal{N}=1$ supergravity. We continue in \S\ref{sec:noscale2} to show how small deformations of the superpotential can perturb such no-scale vacua into metastable de Sitter vacua, with all scalars  stabilised. In particular, we discuss the `no-go' result that such perturbations are not possible unless the unperturbed no-scale vacuum has a third real massless direction. 
We also explain the generalities of the decoupling  mechanism that ensures the stability of $2N-3$ real modes of the Hessian, and we derive the surprisingly simple form of the spectrum of the remaining three modes.  
In \S\ref{sec:examples}, we
present realisations of the mechanism in supergravity and string theory, and discuss the generalities of models in which the 
 no-scale modulus
 appears in the superpotential 
 through a single non-perturbative exponential term.
We also consider the realisation of the mechanism in  type IIB flux vacua, 
and derive a geometric condition on the complex structure field space geometry for 
 minimal ${\cal O} (N^0)$ tuning. Finally, we discuss the realisation of the mechanism in STU supergravity models. 
 We conclude in \S\ref{sec:conclusions}, and list some useful formulae in  appendix \ref{sec:formulae}.

\section{\label{sec:noscale} Unbroken no-scale vacua}

The remarkable properties of no-scale supergravities \cite{Cremmer:1983bf} have  long been appreciated by many authors (for a colloquial review of some of the early developments, see for instance \cite{Nanopoulos:1994as}). In this section, we review some elementary results on moduli stabilisation in unbroken no-scale vacua, and introduce some useful notation.

For the no-scale K\"ahler potential of equation \eqref{eq:K} and the superpotential of equation \eqref{eq:W0},
the scalar potential in equation \eqref{eq:V0} has no-scale solutions with $F_T= K_T W(X^i)$ and $F_i=0$.
For  these vacua,  the potential \eqref{eq:V0} results in vanishing entries in the Hessian for all components involving $T$ or $\bT$:
\be
m^2_{T\bT} = m^2_{TT} = m^2_{Ti} = m^2_{\bT i} =0 \, ,
\ee
where we have adopted the notation $\partial^2_{a b} V = e^K m^2_{ab}$ for the entries of the Hessian matrix. 
This is of course directly related to the vacuum expectation value of $T$ not being fixed in the no-scale solution.
The fields $X^i$ have non-vanishing entries of the Hessian that are given by (see appendix \ref{sec:formulae} for useful formulae),
\bea
m^2_{ i \bj} &=&Z_{ik} \bZ^k_{~\bar \jmath} + |W|^2 K_{i \bj} \, ,\\
m^2_{i  j} &=& 2 Z_{ij} \bW \, ,
\eea
where we have introduced the  K\"ahler invariant and diffeomorphism covariant symmetric  tensor $Z_{ij} = D_i D_j W =
\partial_i F_j +K_i F_j - \Gamma_{ij}^k F_k$, which for $F_i = 0$ reduces to $Z_{ij} = \partial_i F_j$.
For future reference, we note that in this notation, the critical point equations, $\partial_a V= 0$, can be written as \cite{Denef:2004ze},
\be
Z_{ab} \bF^b = 2\bW F_a \, , \label{eq:oldCP}
\ee
and imply  that the no-scale vacuum enforces $Z_{T i} =0$ and $Z_{TT} = 2 W K_{TT}$, 
for $W \neq 0$.

The Hessian matrix \eqref{eq:Hess} for the no-scale system is then given by,
\renewcommand{\arraystretch}{1.5}
\bea
{\cal H} &=&e^K
\left(
\begin{array}{>{\centering\arraybackslash$} p{1cm} <{$} >{\centering\arraybackslash$} p{1cm} <{$}| >{\centering\arraybackslash$} p{1.2cm} <{$}  >{\centering\arraybackslash$} p{1cm} <{$}}
m^2_{T \bT}& \vec{0}^T & m^2_{TT} & \vec{0}^T \\
\vec{0} & m^2_{i \bj} &  \vec{0} & m^2_{ i j}  \\\hline
m^2_{\bT \bT} & \vec{0}^T & m^2_{\bT T} &  \vec{0}^T \\
 \vec{0} & m_{\bi \bj} & \vec{0} &m^2_{\bi j}  \end{array}
\right)  = \cr
\nonumber\\
&=&e^K
\left(
\begin{array}{>{\centering\arraybackslash$} p{.5cm} <{$} >{\centering\arraybackslash$} p{3.5cm} <{$}| >{\centering\arraybackslash$} p{1cm} <{$}  >{\centering\arraybackslash$} p{3.5cm} <{$}}
0& \vec{0}^T & 0 & \vec{0}^T \\
\vec{0} & Z_{ik}\ {\bZ^k}_{\bj } +K_{i \bj}  |W|^2 &  \vec{0} & 2 Z_{ i j} \bW \\\hline
0& \vec{0}^T & 0 &  \vec{0}^T \\
 \vec{0} & 2\bZ_{\bi \bj} W & \vec{0} &\bZ_{\bi \bk}\ {Z^{\bk}}_{j} +K_{\bi j}  |W|^2  \end{array}
\right)  \, .
\label{eq:Hessian_noscale}
\eea
\renewcommand{\arraystretch}{1}
We note in particular that the `off-diagonal' terms, $m^2_{ij}$ are not necessarily small compared to the `diagonal block', $m^2_{i \bj}$.
Since the potential \eqref{eq:V0} is semi-positive definite and vanishes at the no-scale minimum, the eigenvalues of the Hessian matrix must also be semi-positive definite. To see this explicitly, we diagonalise the Hessian matrix. For ease of representation we neglect the trivial $T$ and $\bT$ directions and choose a basis in which the fields are canonically normalised at the critical point, $K_{i \bj}|_{\rm c.p.} =\delta_{i \bj}$.
The complex symmetric matrix $Z_{ij}$ can be Takagi factorized as  $Z = U \Sigma U^T$, where $U$ is a unitary matrix whose columns are orthonormal eigenvectors of $Z\bZ$, $\Sigma = {\rm diag}(\lambda_1,\ldots \lambda_{N-1})$, and the $\lambda_i$ are real and nonnegative, with $\lambda_i^2$ the eigenvalues of $Z\bZ$.
We note that upon performing the $(2N-2) \times (2N-2)$ unitary transformation
\bea
{\cal H} \to {\cal U}^{\dagger} {\cal H} \  {\cal U} \quad \text{with} \quad
{\cal U} =
\left(
\begin{array}{c c}
U  &  0\\
0 & \overline U
\end{array}
\right) , \label{U} \label{eq:unitary1}
\eea
the matrix ${\cal H}$ can be written as
\bea
{\cal H} =
\left(
\begin{array}{c c}
\Sigma^2+\delta_{i\bj} |W|^2 &   2 \overline W \Sigma  \\
 2 W \Sigma  & \Sigma^2+\delta_{\bi j} |W|^2 \end{array}
\right)  \,  .\label{D}
\eea
After rearranging the rows and columns in an obvious way, ${\cal H}$ takes the block diagonal form
\bea
{\cal H} &=&
\left(
\begin{array}{c c c c c}
\lambda_1^2+|W|^2 & 2 \overline W \lambda_1 & 0 &0 \\
2  W \lambda_1 & \lambda_1^2+|W|^2 & 0& 0 \\
0 & 0 & \lambda_2^2+|W|^2 & 2 \overline W \lambda_2  \\
0 & 0& 2  W \lambda_2 & \lambda_2^2+|W|^2 \\
&&&& \ddots
\end{array} \right)  \, . \label{Hess}
\eea
Each $2\times2$ block can be diagonalised by the unitary transformation,
\be
u_{2\ti 2} =\frac 1 {\sqrt 2 }
\left(
\begin{array}{c c}
1 & 1 \\
e^{i\vartheta_W} & - e^{i\vartheta_W}
\end{array}
\right) \, , \label{eq:unitary2}
\ee
where $W = e^{i \vartheta_W} |W|$ and the corresponding eigenvalues of ${\cal H}$ are given by
\be
m_{i\pm}^2 = \lambda_i^2 \pm 2|W| \lambda_i + |W|^2 = \left(\lambda_i \pm |W| \right)^2 \, , \label{eq:m20}
\ee
showing that  the Hessian $\cal  H$ is manifestly semi-positive definite.

In sum, no-scale vacua have $T$ unfixed with a vanishing mass,  a vanishing cosmological constant and a semi-positive definite spectrum for the $X^i$ fields. In this paper, we investigate how small no-scale breaking perturbations to the superpotential, $\delta W(T, X^i)$,  may lift the $T$ modulus and give rise to meta-stable vacua with a positive cosmological constant.

\section{\label{sec:noscale2}Approximate no-scale vacua}

In this section, we discuss how small perturbations to the superpotential can result in potentials admitting metastable de Sitter vacua. The no-go theorem of \cite{Brustein:2004xn} is then relevant: if the supersymmetry breaking $F$-term is completely aligned with  a single field, $T$, with a no-scale K\"ahler potential, then no dS minima are obtainable, independently of the form of the superpotential. Here, we keep the unperturbed form of the K\"ahler potential, \eqref{eq:K}, and only consider superpotential corrections as
\be
W = W_0 (X) + \delta W(T,X^i)\, , 
\ee
which then necessarily have to induce non-vanishing $F$-terms for some of the fields perpendicular to $T$, as discussed in \cite{Kallosh:2014oja}.

We are particularly interested in the  regime in which the  superpotential correction is small compared to $W_0$, such that $|\delta W/W_0| \ll 1$, and we furthermore assume that derivatives of $\delta W(T,X^i)$ are not very large compared to the scale of the perturbation itself. This can for instance be realised if $\delta W(T,X^i)$ is given by the sum of some non-perturbative corrections which are all small compared to $W_0$.\footnote{We do note however that these conditions are not satisfied in some classes of string compactifications: for example, in KKLT vacua $|W_0|$ is tuned to be small compared to the flux scale and $|\delta W/W_0|$ is not small.} In natural units and denoting partial derivatives of the superpotential with subscripts, we then expect,
\be
{\cal O}(\delta W) \sim {\cal O}(\delta W_T) \sim {\cal O}(\delta W_{Ti}) \ll {\cal O}(W_0) \sim {\cal O}( W_{j})\sim {\cal O}( W_{ij}) \, . \label{eq:scaling1}
\ee
This assumed hierarchy for the superpotential terms justifies a perturbative expansion in the no-scale breaking.

For the approximate no-scale solution, we may write the $F$-terms at the minimum as,
\bea
F_T &=& K_T W + \delta W_T \, ,  \\
F_i &=& \epsilon W  f_i \, ,\label{eq:Fterms1}
\eea
where $f_i$ is a unit vector, $||f_i|| = 1$, and we study the perturbative expansion in the small parameter $\epsilon$. 

\subsection{Critical point equations}\label{sec:cp}

The critical point equations in the $T$ and $X^i$ directions determine much of the structure of the approximately no-scale vacua.
The equation $\partial_T V = 0$ is to leading order in a perturbative expansion in $\epsilon$ and $|\delta W/W_0|$ given by
\be
K_{TT} |W_0|^2 \epsilon^2= -\bW_0\, \delta W_{TT}  - \frac{4}{3}K_T{\rm Re}\left( \bW_0\, \delta W_T\right)  \, .
\label{eq:Vt}
\ee
Two properties of this equation are particularly noteworthy: first, while the  $F$-term in the $X^i$ subspace is corrected at \od, derivatives of $\delta W$ -- which we by equation \eqref{eq:scaling1} relate to the magnitude of $|\delta W|$ -- appear first at order $\ep^2$. Thus, consistently with equation \eqref{eq:scaling1} we can realise the scaling,
\be
{\cal O}(\ep^2 \, W_0) \sim{\cal O}(\delta W) \sim {\cal O}(\delta W_T) \sim {\cal O}(\delta W_{Ti}) \ll {\cal O}(W_0) \sim {\cal O}( W_{j})\sim {\cal O}( W_{ij}) \, . \label{eq:scaling}
\ee
Equation \eqref{eq:Vt} then receives subleading corrections at \odc, and the smallness of the F-terms in the $X^i$ subspace assumed in equation \eqref{eq:Fterms1} may be achieved by the tuning of $W_0(X^i)$.  
Second, the reality of the left-hand side of equation \eqref{eq:Vt} enforces that,
\be
{\rm Im}\left( \bW_0 \delta W_{TT} \right) = 0 \, . \label{eq:EOMphase}
\ee
The remaining critical point equation, $\partial_i V=0$, is most illuminatingly phrased as a condition on the supersymmetric masses of the $X^i$ fields, explicitly given by an eigenvalue equation involving $Z_{ij}$. To see this, we note that to leading order,
\be
Z_{Ti} = \delta W_{Ti} +K_i \delta W_T + K_T F_i = K_T F_i + \O(\ep^2)\, ,
\ee
so that the critical point equation $\partial_i V= 0$, which can also be written as $Z_{ia} \bF^a = 2 \bW F_i$ as in equation \eqref{eq:oldCP}, implies that,
\be
Z_{ij} \bF^j  = - \bW F_i \, ,
\label{eq:EOMi}
\ee
up to corrections of \ods.
Contracting the complex conjugate of the equation above with ${Z_{i}}^{\bj}$ and using the equation again, we find to linear  order in $\ep$,
\be
(Z \bZ)_{i}^{~j}F_j = |W|^2 F_i  \, , \label{eq:ZbZ}
\ee
implying that $Z\bZ$ has an eigenvalue equal to $|W|^2$ to this order (in the basis in which $K_{i \bj } = \delta_{ i \bj}$), with $F_i$ being the corresponding eigenvector.

Equation \eqref{eq:ZbZ} directly affects the spectrum of approximately no-scale vacua. From equation \eqref{eq:m20} we find that it implies the existence of one real field in the $X^i$-sector with a vanishing mass at zeroth order in $\ep$. 
Thus,  
out  of all exact no-scale vacua, only the subset with at least three real massless degrees of freedom
  can be perturbatively lifted to de Sitter vacua (or  deformed to non-supersymmetric AdS vacua). 
  
The interpretation of the critical point equation as a condition on the 
 supersymmetric mass spectrum is quite familiar: in  \cite{Denef:2004cf} it was shown that  for critical points at which supersymmetry is spontaneously broken  in the $X^i$-sector, without a dominant contribution from the no-scale field, obey a similar eigenvalue equation  derived from equation \eqref{eq:oldCP}. Consistently,  supersymmetric Minkowski solutions that can be perturbed to non-supersymmetric vacua also have at least one massless direction \cite{Kallosh:2014oja}.

Finally, upon using the critical point equation \eqref{eq:Vt}, we find that the vacuum expectation value of the scalar potential at the critical point is given by,
\bea
e^{-K}V &=& 2 {\rm Re}(K^T \bW_0 \delta W_T) + |F_i|^2 + {\cal O}(\epsilon^4) \nn  \\
&=& \frac{2}{3} K^T {\rm Re}\left( \bW_0 \delta W_T\right) - K^{T T} \bW_0 \delta W_{TT} + {\cal O}(\epsilon^3)  \, ,
\label{eq:V}
\eea
where the reality of the scalar potential is ensured by equation \eqref{eq:EOMphase}.

\subsection{The perturbed mass-matrix}

By introducing a small amount of supersymmetry breaking in the directions perpendicular to $T$, we avoid the no-go theorem of \cite{Brustein:2004xn} and may potentially find metastable de Sitter vacua. Since the additional supersymmetry breaking is small by assumption, a perturbative expansion around the no-scale vacuum is well-motivated. We now show that the Hessian matrix takes a very simple form upon performing such an expansion.

First, we note that the general expression for perturbed \emph{non-degenerate} eigenvalues up to second order is given by,
\be
m^2_{A }\Big|_{\rm tot} = m^2_A \Big|_0 + \delta m^2_A + \sum_{B\neq A} \frac{|\delta m^2_{AB}|^2 }{m^2_A\Big|_0 - m^2_B\Big|_0}  + \ldots \, .
\label{eq:pert}
\ee
This expression is not  directly applicable to the perturbative analysis of the approximate no-scale vacua  as
three real degrees of freedom (Re$(T)$, Im$(T)$ and one real component of the eigenstate of $Z\bZ$ with eigenvalue $|W_0|^2$) are degenerate and massless at zeroth order in no-scale breaking. Taking this degeneracy into account, we will find equation \eqref{eq:pert} useful in determining the structure of the dominant contributions of the Hessian matrix.
 In the following section we describe how 
 $2N-3$ fields are stabilised with positive definite masses $\gtrsim {\cal O}(m_{3/2})$, 
 one real direction generically is lifted at \od, and the Hessian eigenvalues of the remaining two fields take a strikingly simple form.

\subsubsection{Masses for the \texorpdfstring{$X^i$}{} fields}\label{sec:m11}

We begin by considering the spectrum of all fields perpendicular to $T$.

While one real direction in the $X^i$ field space has a vanishing mass at zeroth order (by equations \eqref{eq:ZbZ} and \eqref{eq:m20}, as discussed above), the remaining $X^i$ fields are expected to obtain masses of order ${\rm max}( \lambda_i, W_0)$. The no-scale breaking induces small perturbations to the exact spectrum of these fields, but for $\epsilon  \ll 1$, the corrections are too small to  destabilise these fields. This partial decoupling -- which is a direct consequence of the no-scale structure of the K\"ahler potential -- significantly simplifies the problem of assessing the stability of the perturbed critical point. In principle, these $2N -3$ real degrees of freedom may still affect the general stability of the system by Hessian cross-couplings
that may destabilise lighter fields. We will see however, that the no-scale structure again  makes these contributions negligible, thus ensuring the complete decoupling of all but one $X^i$ field.

We now turn to the remaining, light $X^i$ field and determine its mass and couplings to leading order in the perturbative expansion of equation \eqref{eq:scaling}. 
These expressions are most easily analysed by considering the sub-matrix of the Hessian that involve only the $X^i$ fields, and in a basis in which the fields at the critical point are canonically normalised, i.e.~$K_{i \bj }|_{\rm c.p.} = \delta_{i \bj}$.  
We will see in \S\ref{sec:decoupling} that this explicit neglect of the cross-terms between the $X^i$ fields and the no-scale modulus is perfectly justifiable to \ods. 

The Hessian submatrix involving only the $X^i$ fields is to linear  order in $\epsilon$ given by, 
\bea
\begin{pmatrix}
 m^2_{i\bj} & m^2_{ij}\\
 m^2_{\bi \bj} & m^2_{\bj j}
\end{pmatrix}
&=&
\begin{pmatrix}
Z_{i k} \bZ^k{}_{\bj} + \delta_{i\bj} |W|^2 & 2 Z_{ij} \bW \\
2 \bZ_{\bi \bj} W &\bZ_{\bi \bar k} Z^{\bar k}{}_{ j} + \delta_{\bi  j} |W|^2\\
\end{pmatrix}
+
\begin{pmatrix}
0& (D_i Z_{jk}) \bF^k\\
(\bar D_{\bi} \bZ_{\bj\bar k}) F^{\bar k}&0\\
\end{pmatrix}\nonumber \, .\\
\label{eq:subHess}
\eea
Thus, in comparison with the no-scale Hessian of the $X^i$ field sector of equation  \eqref{eq:Hessian_noscale}, we
note that the 
only structural correction to this sub-Hessian at \od~is given by the second term of  equation \eqref{eq:subHess}.
We are interested in the lightest $X^i$ field,  
which -- as discussed in \S\ref{sec:cp} -- to linear order in $\ep$ corresponds to the eigenmode of $Z\bZ$ with an eigenvector proportional to $F_i$. The corresponding orthonormalised (sub-)Hessian eigenvectors for the modes with eigenvalues $0$ and $4 |W|^2$ are given by,   
\be
v_{1\pm} = \frac{1}{\sqrt 2} \begin{pmatrix} ~~  f^{\bi}\\ \mp \bar f^{ i}\end{pmatrix}\,, \qquad 
\label{eq:vpm}
\ee
where we, as in equation \eqref{eq:Fterms1}, have  introduced the unit vector $f_i = \exp(-i \vartheta_W) F_i/||F_i||$, where again $\vartheta_W = {\rm arg}(W)$.

We now note that Hessian cross-terms 
between the $v_{1-}$ direction and the more massive $X^i$ fields -- that generically are lifted at \odz~-- enter the mass matrix at \od, as $v_{1\pm}$ are eigenvectors of the Hessian to leading order. Thus, from equation \eqref{eq:pert}, we see that they affect the  mass of the lightest $X^i$ field at second order in perturbation theory by $\ods\ll \od$, and may consistently be neglected. This is a crucial property of the metastable vacua with only ${\cal O}(N^0)$ fine tuning that we construct in this paper. 

The Hessian eigenvalues of the lightest $X^i$ field, which we will refer to as the `1' direction, are to linear order in $\ep$  then given by,  
\bea
m^2_{1+} &=& 4 |W|^2 -  {\rm Re}\, (  (D_i Z_{jk}) \bar f^i \bar f^j \bar f^k \bW)  \epsilon  \label{eq:1plus}
\, ,\\
m^2_{1-} &=&  {\rm Re}\, ( (D_i Z_{jk}) \bar f^i \bar f^j \bar f^k \bW)  \epsilon 
\label{eq:m11} \, .
\eea

As a final word of caution, we note that if  
$ {\rm Re}\, ( (D_i Z_{jk}) \bar f^i \bar f^j \bar f^k \bW) $
is of order \od, then 
the structure of the 
$X^i$ sector sub-Hessian is identical to that of the 
unbroken no-scale Hessian to \od, and the critical point equation \eqref{eq:EOMi} again enforces that the mass of the lightest field vanishes. Consequently, in this case the 
lightest $X^i$ field is lifted at \ods, and cross-couplings with heavier $X^i$ fields can then not be neglected in general.
Metastability then requires that 
 ${\cal O}(N)$ elements of the Hessian matrix  may have to be tuned small.

In sum, most of the fields perpendicular to the no-scale field obtain positive definite masses of the order of ${\rm max}(\lambda_i, W_0)$ (cf.~equation \eqref{eq:m20}) and do not develop instabilities in the approximately de Sitter no-scale vacuum. The lightest $X^i$ field is generically lifted at \od, and the detailed form of the squared mass depends on the covariant tensor $D_a Z_{bc}$. It thus seems plausible that only a very modest amount of tuning of the superpotential parameters can result in positive eigenvalues for all the $X^i$ fields. In \S\ref{sec:examples}, we demonstrate that this is indeed true in explicit examples.

\subsubsection{Decoupling and stabilisation of the no-scale direction}\label{sec:decoupling}

To find a metastable de Sitter minimum, we must also stabilise the no-scale modulus $T$. Upon using the critical point equations \eqref{eq:Vt} and \eqref{eq:EOMi} we find that to quadratic order in $\ep$,
\bea
m^2_{T\bT} &=& -\frac{4}{3}K_T {\rm Re}\left( W \delta \bW_{\bT}\right) \, ,  \label{eq:mtbt} \\
m^2_{TT} &=& K^T \bW \delta W_{TTT} - \frac{4}{3} K_T {\rm Re}\left( \bW \delta W_T\right)  \,.
\label{eq:mtt}
\eea
Since these contributions to the mass of the no-scale modulus are of order \ods, and thus quite small compared to other entries of the Hessian matrix, some care must be taken to ensure that cross-couplings do not destabilise the no-scale field. Naively, this would require the fine-tuning of the $2(N-1)$  elements, $m^2_{T i},~m^2_{\bT i}$ of the Hessian, however, we now show that such  fine-tuning is not necessary:
enforcing the critical point equations ensures that none of the cross-terms contribute with any significant destabilising terms.

In \S\ref{sec:noscale} we showed that $m^2_{iT}$ and $m^2_{i \bT}$ vanish in the no-scale vacuum. The correction at \od~is given by,
\bea
m^2_{i \bT} &=& K_{\bT}\left( Z_{ij} \bF^j + \bW F_i \right) + \ods = \ods \, , \label{eq:mibt} \\
m^2_{i T} &=& K_T \left( Z_{ij} \bF^j + \bW F_i \right) + \ods = \ods \, , \label{eq:mit}
\eea
where we used the critical point equation \eqref{eq:EOMi} in the last step of both equations.

In second order perturbation theory, these terms then contribute negatively to the eigenvalues of the (mostly) $T$-field eigenstates at order ${\cal O}\left( \epsilon^4 W_0^4/m^2_{i\pm}\right)$ according to equation \eqref{eq:pert}. Thus, the contribution from the lightest $X^i$ field is expected to enter at \odc, while the other $X^i$ fields contribute at order ${\cal O}(\epsilon^4)$. However, as the leading contribution from $m^2_{TT}$ and $m^2_{T\bT}$ enter at \ods, these potentially destabilising corrections are subleading and do not contribute significantly to the (de-)stabilisation of $T$ (as long as $N\lesssim 1/\epsilon^2$). Indeed, the smallest eigenvalues of the Hessian matrix are to \ods~given by,
\bea
m^2_{T\pm} &=& m^2_{T\bT} \pm |m^2_{TT}|  \nonumber  \\
&=& - \frac{4}{3} K_T {\rm Re}\left( \bW \delta W_T\right) \pm |K^T \bW \delta W_{TTT} - \frac{4}{3} K_T {\rm Re}\left( \bW \delta W_T\right)|   \,.\qquad
\label{eq:mTpm}
\eea
Equation \eqref{eq:mTpm} is one of our main results, and indicates that -- given some  small amount of supersymmetry breaking in the directions perpendicular to $T$ -- all moduli can be stabilised in approximate no-scale compactifications with a very small amount of tuning.

{\bf Let us summarise:}
The Hessian matrix of the approximately no-scale system is schematically given by,
\renewcommand{\arraystretch}{1.2}
\begin{eqnarray*}
{\cal H} &=&e^K
\left(
\begin{array}{ c | c | c | c | c | c }
\cellcolor{yellow!50} m^2_{T \bT}~& m^2_{T\bar 1} &~~~m^2_{T \bj'}~~~& \cellcolor{yellow!50} m^2_{TT} & m^2_{T 1} &~~m^2_{T  j'}~~  \\ \hline
m^2_{ 1 \bT} & \cellcolor{yellow!50}m^2_{1 \bar 1} & m^2_{1 \bj'} & m^2_{1 T} & \cellcolor{yellow!50} m^2_{11} & m^2_{1 j'} \\ \hline
& &\cellcolor{yellow!50} & & &\cellcolor{yellow!50} \\
m^2_{ i' \bT} & m^2_{i' \bar 1} &\cellcolor{yellow!50} m^2_{i'  \bj'} &
m^2_{ i' T} & m^2_{i'  1} & \cellcolor{yellow!50}m^2_{i'  j'} \\
& & \cellcolor{yellow!50}& & &\cellcolor{yellow!50} \\ \hline
\cellcolor{yellow!50} m^2_{\bT\bT} & m^2_{\bT \bar 1} &~~m^2_{\bT  \bj'}~~ &\cellcolor{yellow!50} m^2_{\bT T}~& m^2_{\bT 1} &~~~m^2_{\bT j'}~~~ \\ \hline
m^2_{\bar 1 \bT} & \cellcolor{yellow!50}m^2_{\bar1 \bar 1} & m^2_{\bar 1 \bj'} &
m^2_{ \bar 1 T} &\cellcolor{yellow!50} m^2_{\bar 1  1} & m^2_{\bar 1  j'} \\ \hline
& & \cellcolor{yellow!50}& & & \cellcolor{yellow!50}\\
m^2_{ \bi' \bT} & m^2_{\bi'  \bar 1} &\cellcolor{yellow!50} m^2_{\bi' \bj'} &
m^2_{ \bi' T} & m^2_{\bi'  1} &\cellcolor{yellow!50} m^2_{\bi'  j'} \\
& & \cellcolor{yellow!50}& & & \cellcolor{yellow!50}\\
\end{array}
\right)  =
\end{eqnarray*}
\bea
= m_{3/2}^2
\left(
\begin{array}{ c | c | c | c | c | c }
\cellcolor{yellow!50}\sim \ep^2~& \sim\ep^2~&~~~\sim\ep^2~~~& \cellcolor{yellow!50} \sim\ep^2~& \sim\ep^2 ~&~~\sim\ep^2~~  \\ \hline
\sim \ep^2  & \cellcolor{yellow!50} \sim\ep & \sim \ep & \sim \ep^2 & \cellcolor{yellow!50} \sim \ep & \sim \ep \\ \hline
& &\cellcolor{yellow!50} & & & \cellcolor{yellow!50} \\
\sim \ep^2 & \sim \ep &\cellcolor{yellow!50} \sim 1 &
\sim \ep^2 & \sim \ep &\cellcolor{yellow!50} \sim 1\\
& & \cellcolor{yellow!50} & & &\cellcolor{yellow!50} \\ \hline
\cellcolor{yellow!50}\sim \ep^2  & \sim \ep^2  &~~\sim \ep^2 ~~ &\cellcolor{yellow!50} \sim \ep^2 ~& \sim \ep^2 &~~~\sim \ep^2 ~~~ \\ \hline
\sim \ep^2 &\cellcolor{yellow!50} \sim \ep & \sim \ep  &
\sim \ep^2  &\cellcolor{yellow!50} \sim \ep  & \sim \ep \\ \hline
& &\cellcolor{yellow!50} & & &\cellcolor{yellow!50} \\
\sim \ep^2 & \sim \ep & \cellcolor{yellow!50} \sim 1&
\sim \ep^2 & \sim \ep &\cellcolor{yellow!50} \sim 1\\
& &\cellcolor{yellow!50} & & &\cellcolor{yellow!50} \\
\end{array}
\right)
\, ,
\eea
\renewcommand{\arraystretch}{1}
where we for simplicity of presentation have taken ${\cal O}(\lambda_a) \sim {\cal O}(|W_0|)$ to set the scale of the matrix, and highlighted the dominant elements.

We have shown that most $X^i$ fields decouple at zeroth order in the no-scale breaking and receive squared masses of the order of \odz, cf.~equation \eqref{eq:m20}. The critical point equations imply that only no-scale systems with three real flat directions at zeroth order in the no-scale breaking support approximately de Sitter solutions, as discussed in  \S\ref{sec:cp}. One of these directions (here taken to be the $v_{1-}$-direction) is generically lifted at \od, and the corresponding squared mass may be rendered positive by a modest amount of tuning. The remaining two flat directions are lifted at \ods, with no significant contribution coming from the cross-terms with other moduli.

Thus, the approximately no-scale de Sitter vacua are meta-stable if $m^2_{1-}> 0$, which corresponds to
\be
 {\rm Re}\, ( (D_i Z_{jk}) \bar f^i \bar f^j \bar f^k \bW)  > 0 \, ,
 \ee
 and $m^2_{T-} = m^2_{T \bT} - | m^2_{TT}| >0$, as given by equation \eqref{eq:mTpm}.
 The tuning of the Hessian matrix necessary to obtain such vacua is of order ${\cal O}(N^0)$, which should be compared with ${\cal O}(N^2)$ for similar vacua obtained in \cite{Kallosh:2014oja}.

\section{Examples in supergravity and string theory}
\label{sec:examples}
We expect that the general properties of the approximately no-scale vacua considered in this paper are broadly applicable to many different examples, including those involving more complicated, multi-field no-scale K\"ahler potentials. Here, we will as a proof of principle demonstrate that this type of vacua is indeed obtainable in supergravity, and in particular we focus on string theory inspired supergravities of STU-type. In the simplest scenarios, we find some limiting conditions. For more involved examples, we expect these constraints to be relaxed, and we look forward to exploring these possibilities in future work.

Before delving into the details of an explicit STU-model, we first note some common properties of large classes of models.

\subsection{Models with one non-perturbative effect: generalities}
\label{sec:oneNP}
The simplest model one can consider which realises decoupling from approximate no-scale vacua involves a single non-perturbative, no-scale breaking superpotential correction,
\be\label{eq:deltaW}
\delta W = A(X^i) e^{-a T} \, .
\ee
By performing a trivial, constant K\"ahler transformation, we can choose $W_0 \in \mathbb{R}_+$ at the critical point. The stabilisation of the $T$-axion is ensured by equation \eqref{eq:EOMphase}, from which we infer that at the critical point,
\be
\delta W = s |\delta W| = s |A| e^{-a\tau} \, ,\qquad \tau = {\rm Re}\, T\, ,
\ee
where $\tau$ denotes the real part of $T$ and $s\in\{-1,+1\}$. From equation \eqref{eq:mtbt} we see that $s=+1$ corresponds to $m_{T\bT}<0$, which immediately implies that there is a tachyon. We therefore set $s=-1$. The value of the scalar potential at the critical point is given by equation \eqref{eq:V}, which evaluates to
\bea\label{eq:atbigger1}
V &=& e^K K^{T T} a^2 W_0 |\delta W| \left(1- \frac{1}{a \tau} \right) \, .
\eea
Thus, minima with $V >0$ have $a\tau>1$. The smallest scalar mass is easily computed from equation \eqref{eq:mTpm}
\bea
m^2_{T-} &=& m^2_{T\bT} - |m^2_{TT}|=    \frac{2a  }{\tau} W_0 |\delta W| \left( 1 - |1 - \tau^2  a^2 | \right)\, .
\eea
Thus, we find that $m^2_{T-} > 0$ for $a\tau < \sqrt{2}$.  In sum, metastable de Sitter can be found only for
\be
1< a\tau < \sqrt{2} \, . \label{eq:range}
\ee
The above constraint \eqref{eq:range} appears very restrictive and one might wonder whether it can be realised in a controlled supergravity regime. Since we clearly want $\tau=\text{Re}(T) \gg1$ to suppress $\alpha'$ corrections we need $a\ll1$. In type IIB string theory this may be achievable if the non-perturbative term \eqref{eq:deltaW} arises from gaugino condensation on a large stack of D7-branes. In this case the full correction has the form $\delta W = A e^{-a f(T,X^i)}$ where $f(T,X^i)$ is the gauge kinetic function  \cite{Haack:2006cy}. Due to the non-renormalization theorem of \cite{Antoniadis199137}, the gauge kinetic function takes the form $f(T,X^i) = f^{\rm tree}(T) + f^{\rm 1-loop}(X^i) + f^{\rm np}(T,X^i) = T + f^{\rm 1-loop}(X^i) + \sum_{n=1}^{\infty} c_n(X^i) e^{-anT}$. For Re$(T) \gg 1$, it seems reasonable to assume that $f(T,X^i) \approx f^{\rm tree}(T)=T$, but it would certainly be interesting to check this in explicit models.

Furthermore, the critical point equation \eqref{eq:Vt} implies to leading order in $\epsilon$ that,
\bea
W_0^2 \epsilon^2 &=& \frac{4}{3}W_0 |A|  f(a\tau) \, ,\label{eq:cptau}
\eea
where we have defined $f(x) = (x^2+2x)\exp(-x)$. The function $f(x)$ is semi-monotonically increasing in the range $x\in[1, \sqrt{2}]$, and reaches a maximum at $x=\sqrt{2}$. As $f(x)$ is bounded, this constrains the ratio between $\epsilon^2$ and $|A|/W_0$ obtainable in any model. In particular we find
\be
\frac{3}{4} \frac{\epsilon^2}{f_{\rm max}} < \frac{|A|}{W_0} < \frac{3}{4} \frac{\epsilon^2}{f_{\rm min}} \, ,
\ee
where $f_{\rm min} = f(1) = 3/e \approx 1.104$ and $f_{\rm max} = f(\sqrt{2}) = 2 \left(1+\sqrt{2}\right)\exp(-\sqrt{2})\approx 1.174$. Numerically, we then have,
\be
0.639 < \frac{1}{\ep^2} \frac{|A|}{W_0} < 0.680 \, .\label{eq:A_ratio}
\ee
Thus any realisation of de Sitter vacua through the mechanism we have described above including the exact no-scale K\"ahler potential for a \emph{single} no-scale field and a \emph{single} non-perturbative term is somewhat constrained. As we have seen, the constraint \eqref{eq:A_ratio} follows directly from the critical equation $\partial_T V =0$, which requires $\delta W_T  =  O(\epsilon^2)$, and the metastability condition $a\tau <\sqrt{2}$ from equation \eqref{eq:range}.
 Adding more K\"ahler fields could in principle allow for less restricted models, and it would be interesting to study this in detail.

\subsection{Simple examples in supergravity}

In this section, we illustrate the general mechanism of decoupling by an approximate no-scale structure in the simplest possible supergravity models. While we do not expect these to  directly capture the intricacies of supergravities arising from the dimensional reduction of string theory, they nicely illustrate the basic properties of the general mechanism.

First, we consider a system of two chiral superfields, $T$ and $X$ with the K\"ahler potential,
\be
K = -3 \ln(T+\bT) + X \bX \, ,
\ee
and the superpotential $W= W_0(X) + W_{np}(T)$ with,
\bea
W_0 &=& w_0 + \epsilon w_0 X - \frac{w_0}{2} X^2 + \frac{w_3}{3!} X^3 \, , \\
W_{np} &=& A e^{-a T} \, ,
\eea
where we take $w_0, w_3 \in \mathbb{R}$.
For $\ep\ll1$, this system has a non-supersymmetric solution at $X=0$ with $F_X = \epsilon w_0$ and $F_T = K_T w_0 + {\cal O}(W_{np})$. Consistently with equations \eqref{eq:m20} and \eqref{eq:ZbZ}, the real and imaginary components of $X$ have masses which to zeroth order in $\ep$ are given by $m^2_{X+} = 4w^2_0,$ and $m^2_{X-} = 0$. At linear order in $\ep$ we find that,
\bea
m^2_{X \bX} &=& 2w_0^2 + {\cal O}(\ep^2) \, , \\
m^2_{XX} &=& -2w_0^2 +\ep w_0 w_3 + \ods \, .
\eea
The lightest component of the $X$ field is then metastable for $0<\ep w_0 w_3<4w_0^2$, and the stabilisation of the no-scale modulus may proceed as in \S\ref{sec:oneNP}. Thus, we conclude that the mechanism presented in this paper admits very simple explicit realisations.

The decoupling mechanism discussed in this paper 
 is readily generalised to systems with many interacting fields, and can, in particular be realised in  
 `random supergravity' \cite{Marsh:2011aa}. 
To illustrate this, we consider an ensemble of critical points 
in which the value of the superpotential and the 
 tensors $Z_{ij}$ and $D_i Z_{jk}$ are taken to be random tensors with independent and identically distributed entries, subject to the
  critical point equations \eqref{eq:Vt} and \eqref{eq:EOMi} . The no-scale sector involving $T$ is \emph{not} assumed to be random, but rather to be stabilised as in 
  the discussion of
  \S\ref{sec:oneNP}, or generalisations thereof. 
  Consistently with our general discussion, simulations of such systems with ${\cal O}(100)$ fields show that most of the $X^i$ fields are stabilised with masses $m \gtrsim m_{3/2}$, and a single real field in the $X^i$-sector remains significantly lighter than $m_{3/2}$.
  The fraction of all critical points that are metastable vacua is $P =1/2$, independently of $N$, in agreement with  our discussion around equation \eqref{eq:m11}.
   This should be compared with the
corresponding fractions for critical points  without an underlying no-scale structure \cite{Marsh:2011aa, unpublished, Johannes}: for typical critical points with  $\langle W\rangle^2 \approx {\cal O}({\rm eig}(Z \bZ))$,
a fraction of $P \approx\exp\left(- 0.2 N^2\right)$ of the  typical critical points with  are metastable, and for approximately supersymmetric critical points with $\langle W\rangle^2 \ll {\cal O}({\rm eig}(Z \bZ))$, the metastable fraction is given by $P\approx{\rm exp}(-0.35 N)$.
Thus, in random supergravity, the relative frequency, $P$, of metastable approximately no-scale vacua is exponentially larger than 
the corresponding value for vacua without this structure. 

\subsection{Examples in type IIB string theory: generalities}
\label{sec:GVW}
An appealing feature of the class of de Sitter vacua presented in this paper is that they rely on ingredients which are readily available in compactification of string theory on Calabi-Yau manifolds. In particular, flux compactifications of type IIB string theory include all the ingredients necessary to realise this scenario: the K\"ahler potential for the K\"ahler moduli sector is of the no-scale form to leading order in the $\alpha'$ and
$g_s$ expansions, and it is plausible that among the possible choices of quantised three-form flux,
there are theories admitting $\epsilon \ll 1$. In this section, we consider such flux vacua in more detail, and
derive a geometric condition on the complex structure moduli space which must be satisfied in order to achieve de Sitter vacua with an ${\cal O}(N^0)$~fine-tuning of the Hessian matrix.

Type IIB string theory compactified on an orientifold, $\tilde M_3$, of a Calabi-Yau threefold $M_3$, reduces to a four-dimensional ${\cal N}=1$ effective supergravity if the supersymmetry breaking scale is much smaller than the compactification scale. The relevant degrees of freedom of such compactifications include the axio-dilaton, $S= e^{-\phi}- i C_0 $, the complex structure moduli, $U^i$, where $i=1,\ldots, h^{2,1}_-(\tilde M_3)$, the complexified K\"ahler moduli, $T^r$, where $r=1, \ldots, h^{1,1}_+(\tilde M_3)$, the 2-form axion multiplets $G^\alpha$ with $\alpha = 1,\ldots,h^{1,1}_-(\tilde M_3)$ and possibly open string moduli. Many simple orientifolds have $h^{1,1}_-(\tilde M_3)=0$ and therefore no scalars arising from the $B_2$ and $C_2$ fields. For simplicity, we here consider only this case, and we furthermore  do not explicitly include the action of additional open string degrees of freedom arising from stacks of D-branes.

The leading order K\"ahler potential for the K\"ahler moduli is given by $K_{\rm no-scale}=-2\ln{\cal V}$, where ${\cal V}$ denotes the volume of the compactification manifold in units of $\alpha'$. In this paper, we focus on models with a single K\"ahler modulus, $T$, and as ${\cal V} \sim (T+\bT)^{3/2}$, the K\"ahler modulus may then play the role of the no-scale field of equation \eqref{eq:K}.

The K\"ahler potential for the complex structure moduli, $U^i$, and the axio-dilaton, $S$, is given by
\be
\tilde{K}(S, \bS, U^i, \bar U^{\bi}) = - \ln(S+\bS) - \ln( i \int_{\tilde M_3} \Omega \wedge \bar \Omega ) \, ,
\ee
where $\Omega$ denotes the holomorphic $(3,0)$-form on $\tilde M_3$.
Thus, to leading order in $\alpha'$ and $g_s$, and in the absence of open  string moduli, these fields do not mix with $T$ in the K\"ahler potential and may serve as our $X^i$ field sector.

We discuss the superpotential and its K\"ahler  covariant derivatives, for more details see for instance \cite{Denef:2004ze}. Compactifications with quantised NS-NS flux, $H_3 = dB_2$, and RR flux,  $F_3 = dC_2$, on non-trivial three-cycles of $\tilde M_3$ have a superpotential which is a linear combination of the periods of $\Omega$,
\be
W = \int_{\tilde M_3} \Omega \wedge G_3  = \vec{N} \cdot \vec{\Pi} \, ,
\ee
where we have introduced the complex three-form flux $G_3 = F_3 - i S H_3$,  which takes on $2(h^{2,1}+1)$ quantised values as $\vec N=\vec f - iS \vec h$.  Here $\vec \Pi$ denotes the period vector of the  three-form $\Omega$.

The $F$-terms in the axio-dilaton and complex structure sector are then given by
\bea
&F_S=D_S W =& - \frac{\vec{N}^* \cdot \vec{\Pi}}{S +\bar S}  \, , \label{eq:Fs} \\
&F_i=D_i W =&   \vec{N} \cdot D_i \vec{\Pi} \, .
\eea
The tensor $Z_{ab}$ has the components,
\bea
&Z_{SS} =& 0 \, , \label{eq:Zss}
\\
&Z_{Si} =&   - \frac{\vec{N}^* \cdot D_i \vec{\Pi}}{S + \bar S}  \label{eq:Zsi}\, , \\
&Z_{ij} =&  {\cal F}_{ijk} \vec{N}\cdot  \bar D^{k} \vec{\Pi}^*  = -
(S + \bar S) {\cal F}_{i j k}  \bZ^k_{~\bar S}
\, , \label{eq:Zij}
\eea
where we have used the identity $D_i D_j \Omega = {\cal F}_{i jk } \bar D^{k} \bar \Omega$, with ${\cal F}_{ijk}$ denoting the `Yukawa couplings' of the special geometry.\footnote{The coefficients ${\cal F}_{ijk}$ can be obtained by taking three holomorphic derivatives of the ${\cal N}=2$ prepotential, or by evaluating ${\cal F}_{ijk} = i \int \Omega(U) \wedge  \frac{\partial^3 \Omega(U)}{\partial U^i \partial U^j \partial U^k}= i \int \Omega(U) \wedge  D_i D_jD_k \Omega(U)$.  These `Yukawa couplings'  determine the field space Riemann curvature as, $R_{i \bj k \bar \ell} = - K_{i \bj} K_{k \bar \ell} - K_{i \bar \ell} K_{k \bj} + {\rm exp}(K) {\cal F}_{ik m} \bar {\cal F}_{\bj \bar \ell}^{~~m}$.}

Furthermore, upon acting with three covariant derivatives on the superpotential, we find the components,
\bea
D_S Z_{SS} &=& 0 \, ,~~
D_i Z_{SS} = 0 \, , \\
D_i Z_{jS} &=& - \frac{{\cal F}_{ijk}  \bF^k}{S+ \bS}   \, , \\
D_i Z_{j k} &=&
{\cal F}_{ijk} \vec N \cdot \vec{\Pi}^* +
D_k({\cal F}_{ij\ell}) \vec N \cdot \bar D^\ell \vec{\Pi}^* = \nonumber \\
 &=&
-{\cal F}_{ijk} (S + \bS) \bF_{\bS} -
D_k({\cal F}_{ij\ell})
(S + \bS) \bZ^\ell_{~\bS}
 \, . \label{eq:Uijk}
\eea
Equations \eqref{eq:Fs}--\eqref{eq:Uijk}  determine much of the structure of the type IIB flux vacua.
Of crucial importance for our discussion of the  ${\cal O}(N^0)$ fine-tuning of the Hessian matrix in \S\ref{sec:m11} was that the lightest $X^i$ field was lifted at \od. We noted that if 
 ${\rm Re}\, (\bW  D_i Z_{jk} \bar f^i \bar f^j \bar f^k )$ 
 was of \odz, then cross-couplings with heavier $X^i$ fields were negligible, and meta-stability requires the fine-tuning of ${\cal O}(N^0)$ terms. In contrast, for  ${\rm Re}\, (\bW  D_i Z_{jk} \bar f^i \bar f^j \bar f^k ) \sim \epsilon$ we found an additional cancellation of
the remaining \od~corrections, and the lightest $X^i$ field would be lifted at \ods~with a generic ${\cal O}(N)$ fine-tuning of the Hessian as a result. Here, we note that equation \eqref{eq:Uijk} implies that  the condition of minimal ${\cal O}(N^0)$ fine-tuning of the Hessian matrix in type IIB string theory can be phrased as a geometric constraint on the field space curvature. More precisely, 
minimal tuning  requires:
\be
D_i({\cal F}_{jk\ell})
\bar f^i \bar f^j \bar f^k 
\bZ^\ell_{~\bS} = {\cal O}(\epsilon^0)~~~{\rm [for~}{\cal O}(N^0){\rm ~finetuning]} \, .
\label{eq:condition}
\ee
While this condition is expected to be satisfied for generic `special geometries', it fails in some well-studied models in which the complex structure moduli space is a symmetric space with covariantly constant Riemann curvature for which $D_k({\cal F}_{ij\ell})=0$.
As discussed in \S\ref{sec:m11}, in this case the lightest $X^i$ field is lifted at \ods~together with the no-scale field $T$, and cross-couplings with the heavier complex structure moduli are no longer automatically negligible. However, as we now illustrate in the case of the STU-models, even in this case one can construct examples of this type of vacua.

\subsection{The STU-model with one non-perturbative effect}
\label{sec:STU}
Let us exemplify our findings in the simple `STU-model' with three complex moduli: the axio-dilaton $S=e^{-\phi} - i C_0$, a complex structure modulus $U$ and a complexified volume modulus $T$. This model arises by compactifying type IIB string theory on $T^6/\mathbb{Z}_2 \times \mathbb{Z}_2$, if we restrict ourselves to the isotropic sector and take the three $T^2$ in the $T^6$ to be equal. The latter can be for example imposed by an extra $\mathbb{Z}_3$ symmetry that rotates the three $T^2$, see section 2 in \cite{Shelton:2005cf}. 
In this model, the complex structure moduli space is a symmetric space, so that from our general discussion around the condition \eqref{eq:condition}, we expect $m^2_{1-} \sim \ep^2 m_{3/2}^2$ and that ${\cal O}(N) = {\cal O}(3)$  terms need to be fine-tuned to ensure the metastability of the lightest direction in the $SU$ subspace. 

In the presence of $F_3$ and $H_3$ fluxes the resulting K\"ahler and superpotential are
\bea
K &=&  -3 \log(T+\bT)- \log(S+\bar{S})- 3\log(U+\bar{U})\,,\\
W &=& a_0 + 3 i a_1 U + 3a_2 U^2 + i a_3 U^3 + S(i b_0 +3 b_1 U +3 i b_2 U^2 +b_3 U^3)
\nonumber \\
&+& W_{np}(S, T, U)
\,,
\eea
where the real coefficients $a_i, b_i \in \mathbb{Z}$ correspond to the $F_3$ and $H_3$ fluxes respectively. In particular, if we set $(2\pi)^2 \alpha'=1$, then
\be
\int_{\Sigma_i} F_3 = a_i\,, \qquad \int_{\Sigma_i} H_3 = b_i\,,
\ee
where the $\Sigma_i$ are 3-cycles in the integer homology of the orientifolded space. This leads to $\vec N =\vec f - iS \vec h = \vec a -iS \vec b$ and the above superpotential. The above K\"ahler and superpotential give rise to the usual type IIB no-scale Minkowski vacua discussed in \cite{Giddings:2001yu}.

The volume modulus $T$ appears in the superpotential only through non-per\-tur\-ba\-tive corrections, which arise from Euclidean D3-branes and from gaugino condensation on $D7$-branes, and to leading order take the form $W_{np} = A(S,U) e^{-aT}$. Naturally these non-perturbative corrections are small compared to the tree-level flux contribution so that we can think of them as a perturbation around the no-scale Minkowski vacuum and apply our general approach. Since the moduli $S$ and $U$ are stabilised by the fluxes at tree-level one can assume that the small perturbation does not shift their minimum much so we can neglect the dependence of $A$ on $S$ and $U$ and treat $A$ effectively as a constant:\footnote{As we discussed above there should be one real direction among $S$ and $U$ that gets generically a mass of $\mathcal{O}(\epsilon)$ which is still sufficient to treat $A$ as constant. In the simple STU-model there are further cancellations and this light direction only gets a mass at $\mathcal{O}(\epsilon^2)$ so that the dependence of $A$ on this light direction could be important. It would be interesting to explicitly calculate the function $A(S,U)$ in this model to check whether it is justified to take $A$ to be constant.}
\be
W_{np} = A e^{-a T}\, .
\ee

To simplify the equations, we set $a_1=a_3=b_0=b_2=0$. Then $W$ and all its K\"ahler covariant derivatives are real, if we set the imaginary parts of $S$, $T$ and $U$ to zero. In particular this means that we can trivially solve half of the critical point equations by setting Im$(S)=$Im$(T)=$Im$(U)=0$.

It is convenient to solve this system by considering the `inverse problem' of finding the fluxes which will allow for a minimum  at the moduli vevs  
$S= S_0$, $T= T_0$ and $U=U_0$ with $S_0, T_0,U_0 \in \mathbb{R}$. By appropriate $SL(2,\mathbb{Z})$ transformations we can always ensure that $S_0$ and $U_0$ are in the fundamental domain (i.e.~being larger or equal than one). This is important if we want to count dS vacua, but since we are only interested in their existence we refrain from explicitly doing these $SL(2,\mathbb{Z})$ transformations. Furthermore, our supergravity solutions require $S_0 > 1$ and $T_0 \gg1$ in order to ensure that string loop and $\alpha'$ corrections are small.

The critical point equations 
are solved by the $F$-terms, 
\bea\label{eq:Fterms}
F_S &=& \frac{1}{4 S_0} W_0^{(\ep=0)} \,\ep\,,\cr
F_T &=& K_T W + \frac \lambda {T_0} \ep^2\,,\cr
F_U &=& \frac {3} {4 U_0} W_0^{(\ep=0)} \,\ep\,,
\eea
where $\lambda = T_0 \,\delta W_T/\ep^2$ is a real number of order $\odz$, and where $W_0^{(\ep=0)}$  denotes the flux superpotential at the critical point to {\it zeroth order} in $\ep$, or equivalently, the value of $W$ at the non-supersymmetric  Minkowski minimum with unbroken no-scale symmetry. The flux superpotential can be expressed in terms of $\lambda$ and $T_0$ as, 
\be
W_0^{(\ep=0)} = \frac{4\lambda}{3}(2 +a T_0)\, .
\ee
The choice of normalising  the F-terms of equation  \eqref{eq:Fterms1} in terms of $W_0^{(\ep=0)}$ is convenient as this way the $\epsilon$-expansion of the fluxes truncates  at second order:
\bea
a_0&=& \lambda \lp -\frac{1}{6}\left(7+2 a T_0\right) \epsilon  + \frac{\left(8+a T_0 \left(4+a T_0\right)\right)}{4 a T_0 \left(2+a T_0\right)}\epsilon ^2 \rp\,, \\
a_2&=& \frac \lambda {U_0^2}\left(\frac{2 \left(2+a T_0\right)}{9}+\frac{7 }{18}\epsilon +\frac{ \left(8+a T_0\right)}{36 \left(2+a T_0\right)}\epsilon ^2\right) \,,\\
b_1  &=& \frac{U_0}{S_0}   a_2\,,\\
b_3&=&\frac \lambda {S_0 U_0^3}\left( \frac{  \left(1+2 a T_0\right)}{6}\epsilon+ \frac{a  T_0}{4(2 + a T_0)}\epsilon ^2 \right)\,,\\
A &=& -\epsilon ^2 \frac{e^{a T_0}  \lambda }{a T_0}\,.
\eea
The $F$-terms in the $(S,U)$ subspace of equation \eqref{eq:Fterms} are aligned with the unit vector  $f_i^T = (1/(4S_0), 3/(4 U_0) )$ as in \eqref{eq:Fterms1} and \S\ref{sec:decoupling}, which  upon canonical normalisation is given by $f_i^T = (1/2,  \sqrt{3}/2)$, independently of the field vevs and fluxes. The  gravitino mass at the de Sitter vacuum is 
to \odz~
given by,
\be
m_{3/2}^2 = e^K |W|^2 = \frac{(2+a T_0)^2  \lambda^2}{72 S_0 T_0^3 U_0^3}   \,. 
\ee

We may now proceed as in \S\ref{sec:noscale2} to systematically extract the eigenvalues of the Hessian matrix.

The heaviest fields in the system correspond to the linear combination of $S$ and $U$ that are perpendicular to $f_i$. These fields are lifted by the fluxes at order $\mathcal{O}(\ep^0)$ with the physical masses (we define $M_i^2$ to be the eigenvalues of $V''$ in a canonically normalised basis):
\bea
M^2_{2+} &=&  \frac{16}{9} m_{3/2}^2  \, , \\
M^2_{2-} &=&  \frac{4}{9} m_{3/2}^2  \, ,
\eea
up to corrections of \od.
In addition, one heavy real degree of freedom 
corresponding to the $v_{1+}$ direction of equation 
\eqref{eq:vpm}
is lifted at $\mathcal{O}(\ep^0)$ by the fluxes. By explicit diagonalisation, we  confirm equation \eqref{eq:1plus} to zeroth order in $\ep$,
\be
M^2_{1+} = 4 m_{3/2}^2 \,.
\ee
The remaining three real degrees of freedom corresponding to $v_{1-}$, Re($T$) and Im($T$) are lifted at order $\ep^2$.  From our general discussion in \S\ref{sec:decoupling}, we expect that if the lightest $X^i$ field (the $v_{1-}$ direction) is lifted at \ods, then an ${\cal O}(N)$ fine-tuning is necessary in order to ensure that cross couplings with heavier $X^i$ fields do not destabilise the vacuum. In this STU example however, we find that such cross-terms -- which we in general expect to enter at \od~-- are absent both at  \od~and \ods. Thus, in this particular model, the lightest $X$-field is lighter than the generic expectation, but this still does not result in an increased amount of tuning needed in order to achieve metastability.

The physical masses of the three lightest degrees of freedom are then given by,
\bea
M^2_{L_1} &=&\frac{2 (aT_0)^2  }{(2+a T_0)}m_{3/2}^2 \epsilon ^2\, , \\
M^2_{L_2\pm} &=&  \frac{1}{2 \left(2+a T_0\right){}^2}m_{3/2}^2\epsilon ^2 
\Big(21-2 a^3 T_0^3+14 a T_0  \\
&\pm& \sqrt{4 a T_0 (a T_0+1) (a T_0 (a T_0 (a T_0 (a T_0+7)+27)+26)-33)+217}\Big)\nonumber \, ,\label{eq:M2pm}
\eea
The metastability condition of these de Sitter vacua (which for only $T$ being lifted at \ods~is given by $aT_0< \sqrt{2}$, cf.~equation \eqref{eq:range}), here translates into the marginally stricter condition  $m^2_{L_{2\pm}} >0$, or,
\bea
1< aT_0 < aT_{\star} \approx 1.383 = \sqrt{1.914}\, ,
\eea
where $a T_{\star}$ is given by equation \eqref{eq:aTstar} in the Appendix.

The cosmological constant for these de Sitter vacua is given by,
\be
V = 
\frac{m_{3/2}^2\epsilon ^2}{(2+a T_0)^2} \left( (a T_0 -1)(2+a T_0) -3 \epsilon-\frac{3}{4} \epsilon^2\right)\,.
\ee

We want to point out that our general analysis provides sufficient and necessary conditions only for $\ep \ll 1$ \emph{and} if all quantities scale with $\ep$ in the way we assumed. For larger $\ep$ or at points in moduli space where the scaling is different, it is very well possible that dS vacua still exist and their features can deviate from our predictions. Since for simple models we can find explicit, analytic families of dS vacua it might be possible to map out the entire parameter space of stable dS vacua that are connected to our analytic families. While we leave this interesting task for the future, we give one simple example for the STU-model that shows that
\begin{itemize}
\item stable dS vacua also exist when we move away from the regime of validity of our general ansatz.
\item we can find examples with correctly quantised fluxes.
\item we can choose the flux numbers so small that even in simple models it is possible to satisfy the tadpole condition.
\end{itemize}
In particular, if we take the superpotential to be
\be\label{eq:Winteger}
W = -2+18 U^2+S \left(6U+ 2 U^3\right)-3 e^{- a T}\,,
\ee
then we find stable dS vacua with $S_0 \approx 2.02$, $U_0 \approx .67$ and $aT_0 \approx 1.73$. In particular $aT_0 > \sqrt{2}$ which disagrees with our general prediction. We can also explicitly calculate the F-terms and 
find that $ ||(F_S,0,F_U)|| / 
W\approx 0.72 $ is not small and we cannot view this solution directly as a small $\epsilon$-deformation of a no-scale Minkowski vacuum. Similarly, $|\delta W_T/ W|$ is not small.

The integer fluxes in the superpotential \eqref{eq:Winteger} are $a_0=-2,a_2=6,b_1=2,b_3=2,$ which leads to the tadpole condition (cf. for example \cite{Font:2008vd})
\be
\frac12 \lp a_0 b_3 + 3 a_2 b_1 \rp=\frac12 ( -4+36) = \frac14 N_{O3} = 16\,,
\ee
where the factor $1/2$ in front of the fluxes is due to the fact that the above tadpole condition is derived in the covering space, before taking into account the orientifold projection. We see that we exactly satisfy the tadpole condition in our simple example.

There are additional tadpole constraints due to the presence of O7-planes. The O7-plane charges and tensions can be canceled by D7-branes, which also need to be present to generate the non-perturbative correction to the flux superpotential. It would be very interesting to study this in detail.

\section{\label{sec:conclusions} Conclusions}
We have shown that ${\cal N}=1$ supergravities in which one of the fields has a  K\"ahler potential of no-scale type  admit de Sitter solutions that can be rendered metastable upon the tuning of only two parameters, independently of the number of chiral fields, $N$. 
These vacua arise from small superpotential perturbations to the well-known  non-supersymmetric no-scale Minkowski vacuum.  The perturbation induces small F-terms in the directions perpendicular to the no-scale modulus, $F_i = {\cal O}(\epsilon W)$. No-scale vacua that can be perturbatively lifted to de Sitter space 
have one real massless mode along the $F_i$ direction in field space, in addition to the massless no-scale modulus. This massless mode appears as a direct consequence of the critical point equation and extends the theorem of \cite{Kallosh:2014oja} for non-supersymmetric deformations of supersymmetric Minkowski solutions to have at least one real flat direction.


At the de Sitter critical point, the lingering approximate no-scale structure of the Hessian matrix ensures that $2N-3$ real modes  receive positive definite eigenvalues that are (at least) of the order of the gravitino mass, 
$m_{3/2}  = e^{K/2} |W|$.
The remaining three modes, which we denote  `$v_{1-}$' and `$T\pm$',  are lifted at ${\cal O}(\ep)$ and ${\cal O}(\ep^2)$, respectively. In general, the stabilisation of such light modes require fine-tuning of off-diagonal Hessian cross-terms with heavier modes, however, in \S\ref{sec:decoupling} we found that  the no-scale structure again leads to a cancellation of the leading order cross-terms, thus reducing the question of the full stabilisation of the vacuum to that of two $2\times2$ matrices.  Metastability can then be obtained by tuning the ratio of the off-diagonal entries to the diagonal entries of these matrices, which we showed corresponds to requiring
\be
 {\rm Re}\, (\bW D_i Z_{jk} \bar f^i \bar f^j \bar f^k ) > 0 \, ,
 \ee
 and ensuring that 
 \be
  - \frac{4}{3} K_T {\rm Re}\left( \bW \delta W_T\right) - |K^T \bW \delta W_{TTT} - \frac{4}{3} K_T {\rm Re}\left( \bW \delta W_T\right)| > 0\, .
  \ee 
With this moderate tuning, the spectrum of the lightest modes is given by,
 \be
 m^2_{1-}   \sim m^2_{3/2}\epsilon >0 
 \,,\qquad 
 m^2_{T\pm}  \sim  m_{3/2}^2  \epsilon^2 >0 
\,.\label{eq:squaredmasses}
\ee
We stress that this mechanism for the decoupling of many modes is different from the well-known supersymmetric  decoupling, in which many fields are given positive definite supersymmetric squared masses at some scale $m^2_{\rm susy} \gg m^2_{3/2}$.  In contrast,  the `decoupling by no-scale' developed in this paper does not require large (or small) supersymmetric masses for any of the fields, but is rather ensured by the very particular no-scale structure. For approximate no-scale de Sitter vacua, the (classical) vacuum energy is hierarchically smaller than the generic supergravity expectation, 
\be
\langle V \rangle \sim m_{3/2}^2 M_{\rm  Pl}^2 \epsilon^2 \ll m_{3/2}^2 M_{\rm Pl}^2 \, .
\ee

Since the no-scale form of the K\"ahler potential often appears in the four dimensional effective theories derived from dimensional reduction of  string theory, we expect that the general scenario presented in this paper is relevant for constructing metastable de Sitter vacua in string theory. In \S\ref{sec:GVW}, we considered the embedding of this scenario in the type IIB `landscape' of flux compactifications, 
in which fine-tuning of three-form fluxes can ensure that  $\ep\ll1$. 
We derived a geometric condition on the complex structure field space geometry for the realisation of the minimal version of this scenario with ${\cal O}(N^0)$ fine-tuning.  
We expect this condition to be satisfied for generic Calabi-Yau compactifications, but we note that it fails in 
 toroidal orientifold compactifications in which the  complex structure moduli space is a symmetric space. 
However, also in this case metastable de Sitter minima can be found, and in \S\ref{sec:STU}
we explicitly constructed such solutions in STU-supergravity models.  

The simplest examples of the mechanism for obtaining metastable de Sitter vacua presented in this paper have two obvious short-comings. First, if the single no-scale modulus considered in this paper is stabilised by a single non-perturbative effect with $\delta W = A \exp\left( -a T \right)$, then metastable de Sitter solutions are only obtainable for $1<a {\rm Re}(T) < \sqrt{2}$ and $0.639 \ep^2 <|A/W_0|< 0.680 \ep^2$. While gaugino condensation on large rank gauge groups gives $a\ll1$, so that ${\cal V} \sim ({\rm Re}(T))^{3/2} \gg 1$ and $m_{3/2} \ll M_{\rm Pl}$ are consistently achievable, 
it remains to be established that  the  compactification volumes can be 
made large enough to  neglect all $\alpha'$ corrections. Moreover, $A$ is in general a modulus dependent function, and we have not fully assessed the severity of the constraint for $|A/W_0|$ in string theory models. 
Second, the mass of the lightest two moduli is comparable to the (classical) vacuum Hubble parameter, 
and this may  complicate the construction of  viable  cosmologies for these vacua.    

However, 
 while in this paper we have considered {\it single no-scale modulus} theories
 and in particular focused on the special case in which 
   the no-scale symmetry is broken by  a \emph{single non-perturbative superpotential correction}, we know of no reason why the general mechanism presented in this paper should not be extendable to more general theories  in which these shortcomings may be overcome.  
 For example, it will be interesting to embed this mechanism in the Large Volume Scenario, 
 which has a multiple moduli no-scale sector and 
 in which the no-scale symmetry is broken by perturbative $\alpha'$ corrections and non-perturbative superpotential corrections, so that exponentially large volumes are obtainable  \cite{Balasubramanian:2005zx, Conlon:2005ki}. 
 There are good reasons to believe that such embeddings 
 can successfully be achieved: small K\"ahler deformations of  no-scale models give a positive diagonal term in the Hessian along the sGoldstino direction \cite{Covi:2008ea}, 
and 
the approximately no-scale
`K\"ahler  uplifting' scenarios such as those outlined in \cite{Westphal:2006tn,Rummel:2011cd}  exhibit a similar mass scaling for the no-scale field  (quadratic in the perturbation) as in our solutions  \eqref{eq:squaredmasses}.


In sum, we anticipate that the methods and results presented in this paper can be usefully applied to find fully controlled, simple  de Sitter vacua in string theory. The explicit construction of such vacua is an important problem for the future.

\acknowledgments
We would like to thank J.~Conlon, A.~Guarino, A.~Hebecker, S.~Kachru, R.~Kallosh, S.~Krippendorf, A.~Linde, A.~Lukas, L.~McAllister, X.~de la Ossa, D.~Roest, M.~Rummel, G.~Shiu and  A.~Westphal	 for illuminating discussions.
We are particularly grateful to R.~Kallosh and S.~Krippendorf for comments on a draft of this paper. 
 D.M. is funded by the European Research Council starting grant ``Supersymmetry Breaking in String Theory'', but the contents of this paper reflect only the authors views and not necessarily the views of the European Commission. B.V. is supported by the European Commission through the Marie Curie Intra-European fellowship 328652--QM--sing. T.W. is supported by a Research Fellowship (Grant number WR 166/1-1) of the German Research Foundation (DFG). DM is very grateful to Birzeit University for kind hospitality while finishing this paper.

\appendix

\section{Useful formulae}\label{sec:formulae}

We often use subscripts to indicate scalar derivatives, for example $\partial_a V \equiv V_a$. The potential and its derivatives at a critical point with  $V_a=0$ are \cite{Denef:2004cf}:
\begin{eqnarray}\label{potential}
V&=&e^{K}\Big(F_a \bar  F^a-3|W|^{2}\Big)\ ,\\
V_a&=& e^K\lp (D_a D_b W )\bar F^b  - 2 F_a \bW\rp =e^K\lp Z_{ab}\bF^b  - 2 F_a \bW\rp \label{eq:dV}\,,\\
V_{ab}&=&e^{K}\Big((D_{a}D_{b}D_{c}W)\bar F^{c}-(D_{a}D_{b}W)\bW\Big) = e^{K}\Big(U_{abc}\bar F^{c}-Z_{ab}\bW\Big)\ ,\label{DDV}\\
V_{a\bar b}&=&e^{K}\Big(-{R}_{a\bar{b}c\bar{d}}\bF^{c}F^{\bar{d}}+K_{a \bar b}F_{c}
\bF^{c}-{F}_{a}\bar F_{\bar b}
+(D_{a}D_{c}W)(\bar{D}_{\bar{b}}\bar{D}^{c}\bW)-2K_{a\bar{b}}|W|^{2}\cr
&=&e^{K}\Big(-{R}_{a\bar{b}c\bar{d}}\bF^{c}F^{\bar{d}}+K_{a \bar b}F_{c}
\bF^{c}-{F}_{a}\bar F_{\bar b}
+(Z \bar{Z})_{a\bar{b}}-2K_{a\bar{b}}|W|^{2}\,,
\end{eqnarray}
where we used the definitions $F_a = D_a W$, $Z_{ab} = D_a F_b$ and $U_{abc} = D_a Z_{bc}$ and $D_a$ is the K\"ahler and diffeomorphism covariant derivative such that for example $Z_{ab} =\partial_a F_b + K_a F_b -\Gamma_{ab}^c F_c$.

The derivatives of our K\"ahler potential with respect to the no-scale modulus are:
\allowdisplaybreaks
\begin{align}
K &= - 3 \ln (T + \bT) + \tilde K(X^i \bX^i) \,, \\
\frac{1}{3}K_T &= \frac{1}{K^T} \, ,\\
K_{T \bT} &= \frac{1}{3} K_T \bar K_{\bT}\,, \\
K_T K_{T \bT} &= \frac{3}{2} K_{T \bT T}\,, \\
\Gamma^T_{TT} &= \frac{2}{3} K_T\,, \\
R_{T \bT T \bT} &= \frac{2}{3} K_{T \bT} K_{T \bT}\,.
\end{align}
In our conventions the K\"ahler potential is symmetric under $T$ and $\bT$ exchange so that derivatives with respect to $T$ and $\bT$ are the same, for example $K_{TT} = K_{T\bT} = K_{\bT\bT}$.

Straightforward calculations lead to the following covariant derivatives:
\begin{align}
W &= W_0(X^i) + \delta W(X^i,T)\,, \\
F_T &= K_T W + \delta W_T\,, \\
F_i &= K_i W + W_i\,,\\
Z_{TT} &= \delta W_{TT} + \frac{4}{3} K_T \delta W_T + 2 K_{TT} W\,, \\
Z_{Ti} &= \delta W_{Ti} + K_i \delta W_T + K_T F_i\,, \\
Z_{ij} &= W_{ij} + K_i F_j + K_j F_i - \Gamma_{ij}^k F_k + (K_{ij}-K_i K_j) W\,,\\
U_{TTT} &=K_{TTT} W+ \delta W_{TTT} + K_T \delta W_{TT} + 2 K_{TT} \delta W_T\,,\\
U_{TTi} &= 2K_{TT} F_i + \partial^2_{TT} F_i + \frac{4}{3} K_T \partial_T F_i\cr
&=2 K_{TT} (K_i W+ W_i) +K_i \delta W_{TT} +  \delta W_{TTi} +\frac43 K_T (K_i \delta W_T+\delta W_{Ti})\,, \\
U_{ij T} &=  (\partial_T + K_T) Z_{ij} = K_T Z_{ij}+ \delta W_{Tij} + K_i \delta W_{Tj} + K_j \delta W_{Ti} -\Gamma_{ij}^k \delta W_{Tk}  \cr
&\qquad\qquad\qquad\qquad + (K_{ij} + K_i K_j-\Gamma^k_{ij} K_k) \delta W_{T}\,.\qquad
\end{align}

\subsection{STU-model}
The requirement $M^2_{L_2-} > 0$ for the squared masses of \eqref{eq:M2pm} determines the  upper bound $a T_0 <  aT_\star$ with, 
\bea
aT_{\star} &=& \frac{1}{12} \left(-12+\sqrt{3 \left(38+c\right)}+ \left[228-3c +576 \sqrt{\frac{3}{38+c}}\right]^{\frac{1}{2}}\right)\approx 1.383 \,,\qquad\label{eq:aTstar}\\
&&c=\sqrt[3]{53477-108 \sqrt{105389}}+\sqrt[3]{53477+108 \sqrt{105389}}\,.\nn
\eea


\bibliographystyle{JHEP}
\bibliography{refs}

\providecommand{\href}[2]{#2}\begingroup\raggedright\begin{thebibliography}{10}

\bibitem{Riess:1998cb}
{\bf Supernova Search Team} Collaboration, A.~G. Riess et~al., {\it
  {Observational evidence from supernovae for an accelerating universe and a
  cosmological constant}},  {\em Astron.J.} {\bf 116} (1998) 1009--1038,
  [\href{http://xxx.lanl.gov/abs/astro-ph/9805201}{{\tt astro-ph/9805201}}].

\bibitem{Perlmutter:1998np}
{\bf Supernova Cosmology Project} Collaboration, S.~Perlmutter et~al., {\it
  {Measurements of Omega and Lambda from 42 high redshift supernovae}},  {\em
  Astrophys.J.} {\bf 517} (1999) 565--586,
  [\href{http://xxx.lanl.gov/abs/astro-ph/9812133}{{\tt astro-ph/9812133}}].

\bibitem{Union2}
N.~{Suzuki}, D.~{Rubin}, C.~{Lidman}, G.~{Aldering}, {Amanullah}, et~al., {\it
  {The Hubble Space Telescope Cluster Supernova Survey. V. Improving the
  Dark-energy Constraints above $z > 1$ and Building an Early-type-hosted
  Supernova Sample}},  {\em \apj} {\bf 746} (Feb., 2012) 85,
  [\href{http://xxx.lanl.gov/abs/1105.3470}{{\tt arXiv:1105.3470}}].

\bibitem{Ade:2013zuv}
{\bf Planck Collaboration} Collaboration, P.~Ade et~al., {\it {Planck 2013
  results. XVI. Cosmological parameters}},  {\em Astron.Astrophys.} (2014)
  [\href{http://xxx.lanl.gov/abs/1303.5076}{{\tt arXiv:1303.5076}}].

\bibitem{Kachru:2003aw}
S.~Kachru, R.~Kallosh, A.~D. Linde, and S.~P. Trivedi, {\it {De Sitter vacua in
  string theory}},  {\em Phys.Rev.} {\bf D68} (2003) 046005,
  [\href{http://xxx.lanl.gov/abs/hep-th/0301240}{{\tt hep-th/0301240}}].

\bibitem{Balasubramanian:2005zx}
V.~Balasubramanian, P.~Berglund, J.~P. Conlon, and F.~Quevedo, {\it
  {Systematics of moduli stabilisation in Calabi-Yau flux compactifications}},
  {\em JHEP} {\bf 0503} (2005) 007,
  [\href{http://xxx.lanl.gov/abs/hep-th/0502058}{{\tt hep-th/0502058}}].

\bibitem{Conlon:2005ki}
J.~P. Conlon, F.~Quevedo, and K.~Suruliz, {\it {Large-volume flux
  compactifications: Moduli spectrum and D3/D7 soft supersymmetry breaking}},
  {\em JHEP} {\bf 0508} (2005) 007,
  [\href{http://xxx.lanl.gov/abs/hep-th/0505076}{{\tt hep-th/0505076}}].

\bibitem{Sousa:2014qza}
K.~Sousa and P.~Ortiz, {\it {Perturbative Stability along the Supersymmetric
  Directions of the Landscape}},  \href{http://xxx.lanl.gov/abs/1408.6521}{{\tt
  arXiv:1408.6521}}.

\bibitem{Danielsson:2011au}
U.~H. Danielsson, S.~S. Haque, P.~Koerber, G.~Shiu, T.~Van~Riet, and T.~Wrase,
  {\it {De Sitter hunting in a classical landscape}},  {\em Fortsch.Phys.} {\bf
  59} (2011) 897--933, [\href{http://xxx.lanl.gov/abs/1103.4858}{{\tt
  arXiv:1103.4858}}].

\bibitem{Rummel:2011cd}
M.~Rummel and A.~Westphal, {\it {A sufficient condition for de Sitter vacua in
  type IIB string theory}},  {\em JHEP} {\bf 1201} (2012) 020,
  [\href{http://xxx.lanl.gov/abs/1107.2115}{{\tt arXiv:1107.2115}}].

\bibitem{Cicoli:2012fh}
M.~Cicoli, A.~Maharana, F.~Quevedo, and C.~Burgess, {\it {De Sitter String
  Vacua from Dilaton-dependent Non-perturbative Effects}},  {\em JHEP} {\bf
  1206} (2012) 011, [\href{http://xxx.lanl.gov/abs/1203.1750}{{\tt
  arXiv:1203.1750}}].

\bibitem{Cicoli:2012vw}
M.~Cicoli, S.~Krippendorf, C.~Mayrhofer, F.~Quevedo, and R.~Valandro, {\it
  {D-Branes at del Pezzo Singularities: Global Embedding and Moduli
  Stabilisation}},  {\em JHEP} {\bf 1209} (2012) 019,
  [\href{http://xxx.lanl.gov/abs/1206.5237}{{\tt arXiv:1206.5237}}].

\bibitem{Hebecker:2012aw}
A.~Hebecker, S.~C. Kraus, M.~Kuntzler, D.~L{\"u}st, and T.~Weigand, {\it
  {Fluxbranes: Moduli Stabilisation and Inflation}},  {\em JHEP} {\bf 1301}
  (2013) 095, [\href{http://xxx.lanl.gov/abs/1207.2766}{{\tt
  arXiv:1207.2766}}].

\bibitem{Louis:2012nb}
J.~Louis, M.~Rummel, R.~Valandro, and A.~Westphal, {\it {Building an explicit
  de Sitter}},  {\em JHEP} {\bf 1210} (2012) 163,
  [\href{http://xxx.lanl.gov/abs/1208.3208}{{\tt arXiv:1208.3208}}].

\bibitem{Danielsson:2012by}
U.~Danielsson and G.~Dibitetto, {\it {On the distribution of stable de Sitter
  vacua}},  {\em JHEP} {\bf 1303} (2013) 018,
  [\href{http://xxx.lanl.gov/abs/1212.4984}{{\tt arXiv:1212.4984}}].

\bibitem{Blaback:2013ht}
J.~Bl{\aa}b{\"a}ck, U.~Danielsson, and G.~Dibitetto, {\it {Fully stable dS
  vacua from generalised fluxes}},  {\em JHEP} {\bf 1308} (2013) 054,
  [\href{http://xxx.lanl.gov/abs/1301.7073}{{\tt arXiv:1301.7073}}].

\bibitem{Damian:2013dq}
C.~Damian, L.~R. Diaz-Barron, O.~Loaiza-Brito, and M.~Sabido, {\it {Slow-Roll
  Inflation in Non-geometric Flux Compactification}},  {\em JHEP} {\bf 1306}
  (2013) 109, [\href{http://xxx.lanl.gov/abs/1302.0529}{{\tt
  arXiv:1302.0529}}].

\bibitem{Damian:2013dwa}
C.~Damian and O.~Loaiza-Brito, {\it {More stable de Sitter vacua from S-dual
  nongeometric fluxes}},  {\em Phys.Rev.} {\bf D88} (2013), no.~4 046008,
  [\href{http://xxx.lanl.gov/abs/1304.0792}{{\tt arXiv:1304.0792}}].

\bibitem{Cicoli:2013cha}
M.~Cicoli, D.~Klevers, S.~Krippendorf, C.~Mayrhofer, F.~Quevedo, et~al., {\it
  {Explicit de Sitter Flux Vacua for Global String Models with Chiral Matter}},
   {\em JHEP} {\bf 1405} (2014) 001,
  [\href{http://xxx.lanl.gov/abs/1312.0014}{{\tt arXiv:1312.0014}}].

\bibitem{Blaback:2013qza}
J.~Bl{\aa}b{\"a}ck, D.~Roest, and I.~Zavala, {\it {De Sitter Vacua from
  Non-perturbative Flux Compactifications}},  {\em Phys.Rev.} {\bf D90} (2014)
  024065, [\href{http://xxx.lanl.gov/abs/1312.5328}{{\tt arXiv:1312.5328}}].

\bibitem{Danielsson:2013rza}
U.~Danielsson and G.~Dibitetto, {\it {An alternative to anti-branes and
  O-planes?}},  {\em JHEP} {\bf 1405} (2014) 013,
  [\href{http://xxx.lanl.gov/abs/1312.5331}{{\tt arXiv:1312.5331}}].

\bibitem{Rummel:2014raa}
M.~Rummel and Y.~Sumitomo, {\it {De Sitter Vacua from a D-term Generated
  Racetrack Uplift}},  \href{http://xxx.lanl.gov/abs/1407.7580}{{\tt
  arXiv:1407.7580}}.

\bibitem{Ferrara:2014kva}
S.~Ferrara, R.~Kallosh, and A.~Linde, {\it {Cosmology with Nilpotent
  Superfields}},  {\em JHEP} {\bf 1410} (2014) 143,
  [\href{http://xxx.lanl.gov/abs/1408.4096}{{\tt arXiv:1408.4096}}].

\bibitem{Kallosh:2014via}
R.~Kallosh and A.~Linde, {\it {Inflation and Uplifting with Nilpotent
  Superfields}},  \href{http://xxx.lanl.gov/abs/1408.5950}{{\tt
  arXiv:1408.5950}}.

\bibitem{Kallosh:2014wsa}
R.~Kallosh and T.~Wrase, {\it {Emergence of Spontaneously Broken Supersymmetry
  on an Anti-D3-Brane in KKLT dS Vacua}},
  \href{http://xxx.lanl.gov/abs/1411.1121}{{\tt arXiv:1411.1121}}.

\bibitem{Giddings:2001yu}
S.~B. Giddings, S.~Kachru, and J.~Polchinski, {\it {Hierarchies from fluxes in
  string compactifications}},  {\em Phys.Rev.} {\bf D66} (2002) 106006,
  [\href{http://xxx.lanl.gov/abs/hep-th/0105097}{{\tt hep-th/0105097}}].

\bibitem{Saltman:2004sn}
A.~Saltman and E.~Silverstein, {\it {The Scaling of the no scale potential and
  de Sitter model building}},  {\em JHEP} {\bf 0411} (2004) 066,
  [\href{http://xxx.lanl.gov/abs/hep-th/0402135}{{\tt hep-th/0402135}}].

\bibitem{Kallosh:2014oja}
R.~Kallosh, A.~Linde, B.~Vercnocke, and T.~Wrase, {\it {Analytic Classes of
  Metastable de Sitter Vacua}},  {\em JHEP} {\bf 1410} (2014) 11,
  [\href{http://xxx.lanl.gov/abs/1406.4866}{{\tt arXiv:1406.4866}}].

\bibitem{Marsh:2011aa}
D.~Marsh, L.~McAllister, and T.~Wrase, {\it {The Wasteland of Random
  Supergravities}},  {\em JHEP} {\bf 1203} (2012) 102,
  [\href{http://xxx.lanl.gov/abs/1112.3034}{{\tt arXiv:1112.3034}}].

\bibitem{Cremmer:1983bf}
E.~Cremmer, S.~Ferrara, C.~Kounnas, and D.~V. Nanopoulos, {\it {Naturally
  Vanishing Cosmological Constant in N=1 Supergravity}},  {\em Phys.Lett.} {\bf
  B133} (1983) 61.

\bibitem{Nanopoulos:1994as}
D.~V. Nanopoulos, {\it {The March towards no scale supergravity}},  {\em NATO
  Adv.Study Inst.Ser.B} {\bf 352} (1996) 677--693,
  [\href{http://xxx.lanl.gov/abs/hep-ph/9411281}{{\tt hep-ph/9411281}}].

\bibitem{Denef:2004ze}
F.~Denef and M.~R. Douglas, {\it {Distributions of flux vacua}},  {\em JHEP}
  {\bf 0405} (2004) 072, [\href{http://xxx.lanl.gov/abs/hep-th/0404116}{{\tt
  hep-th/0404116}}].

\bibitem{Brustein:2004xn}
R.~Brustein and S.~P. de~Alwis, {\it {Moduli potentials in string
  compactifications with fluxes: Mapping the discretuum}},  {\em Phys.Rev.}
  {\bf D69} (2004) 126006, [\href{http://xxx.lanl.gov/abs/hep-th/0402088}{{\tt
  hep-th/0402088}}].

\bibitem{Denef:2004cf}
F.~Denef and M.~R. Douglas, {\it {Distributions of nonsupersymmetric flux
  vacua}},  {\em JHEP} {\bf 0503} (2005) 061,
  [\href{http://xxx.lanl.gov/abs/hep-th/0411183}{{\tt hep-th/0411183}}].

\bibitem{Haack:2006cy}
M.~Haack, D.~Krefl, D.~Lust, A.~Van~Proeyen, and M.~Zagermann, {\it {Gaugino
  Condensates and D-terms from D7-branes}},  {\em JHEP} {\bf 0701} (2007) 078,
  [\href{http://xxx.lanl.gov/abs/hep-th/0609211}{{\tt hep-th/0609211}}].

\bibitem{Antoniadis199137}
I.~Antoniadis, K.~Narain, and T.~Taylor, {\it Higher genus string corrections
  to gauge couplings},  {\em Physics Letters B} {\bf 267} (1991), no.~1 37 --
  45.

\bibitem{unpublished}
T.~Bachlechner, M.~Marsh, L.~McAllister, and T.~Wrase, {\it {Unpublished
  notes}}, .

\bibitem{Johannes}
J.~Bausch, {\it {On the Efficient Calculation of a Linear Combination of
  Chi-Square Random Variables with an Application in Counting String Vacua}},
  {\em J.Phys.} {\bf A46} (2013) 505202,
  [\href{http://xxx.lanl.gov/abs/1208.2691}{{\tt arXiv:1208.2691}}].

\bibitem{Shelton:2005cf}
J.~Shelton, W.~Taylor, and B.~Wecht, {\it {Nongeometric flux
  compactifications}},  {\em JHEP} {\bf 0510} (2005) 085,
  [\href{http://xxx.lanl.gov/abs/hep-th/0508133}{{\tt hep-th/0508133}}].

\bibitem{Font:2008vd}
A.~Font, A.~Guarino, and J.~M. Moreno, {\it {Algebras and non-geometric flux
  vacua}},  {\em JHEP} {\bf 0812} (2008) 050,
  [\href{http://xxx.lanl.gov/abs/0809.3748}{{\tt arXiv:0809.3748}}].

\bibitem{Covi:2008ea}
L.~Covi, M.~Gomez-Reino, C.~Gross, J.~Louis, G.~A. Palma, et~al., {\it {de
  Sitter vacua in no-scale supergravities and Calabi-Yau string models}},  {\em
  JHEP} {\bf 0806} (2008) 057, [\href{http://xxx.lanl.gov/abs/0804.1073}{{\tt
  arXiv:0804.1073}}].

\bibitem{Westphal:2006tn}
A.~Westphal, {\it {de Sitter string vacua from Kahler uplifting}},  {\em JHEP}
  {\bf 0703} (2007) 102, [\href{http://xxx.lanl.gov/abs/hep-th/0611332}{{\tt
  hep-th/0611332}}].

\end{thebibliography}\endgroup

\end{document}